\documentclass[aps,pre,preprint]{revtex4}
\usepackage{graphicx}

\begin{document}
\title{Resonance Effects in the Interaction of NLS Solitons with Potential Wells}

\author{ K.T. Stoychev, M.T. Primatarowa and R.S. Kamburova}
\address{ Institute of Solid State Physics\\
 Bulgarian Academy of Sciences, 1784 Sofia, Bulgaria }

\begin{abstract}

The interaction of nonlinear Schr\"{o}dinger (NLS) solitons with potential wells with
variable shapes is investigated numerically. For fixed initial velocities below the
threshold for transmission, the outcome pattern as a function of the width of the
potential yields periodically repeating regions of trapping, transmission and
reflection. The observed effects are explained by an excitation and a following resonant
deexcitation of amplitude (shape) oscillations of the solitons at the boundaries of the
well, associated with radiation modes.

\end{abstract}


\maketitle

Self-localized nonlinear waves (solitons) have been studied in many areas of physics
including optics, solid state, molecular, plasma, elementary particles etc. Within
integrable models, solitons exhibit remarkable stability - they propagate with constant
velocities and shapes and emerge from collisions unchanged except for phase and space
shifts. Real physical systems are often described by nonintegrable equations or such
containing nonintegrable perturbations. This leads to inelastic soliton interactions
with a variety of outcomes. As solitons provide an important mechanism for energy and
information transport in nonlinear systems, such interactions have attracted
considerable attention (see i.e. Ref.~1 for a review of earlier works  on
soliton dynamics in nearly integrable systems). Investigations have been focussed on
collisions between solitons in nonintegrable models and interactions of solitons with
defects and inhomogeneities. In both cases, due to the inelasticity of the interactions,
solitons can change their velocities, break into a number of localized and dispersive
waves and/or be trapped into bound states. In addition, interesting resonance phenomena
have been observed.

Resonace effects in kink-antikink collisions have been studied numerically in some
nonintegrable equations including $\phi^4$, double and modified sine-Gordon and others
\cite{abl,camp1,peyr}. For initial velocities below the threshold for trapping, a
sequence of narrow regions of reflection have been obtained. These reflection windows
have been explained by a "two-bounce" resonance mechanism involving excitation of an
internal shape mode during the first collision, temporal trapping of the solitons due to
loss of kinetic energy, deexcitation of the shape mode during the second (backward)
collision and escape of the kinks to infinity (reflection). The resonance condition
requires that the time between the two collisions is commensurate with the period of the
shape mode. Fine three- and four-bounce resonance structures have also been obtained
\cite{camp2}. Collisions of vector NLS solitons have been investigated in \cite{Yang1}
where fractal resonant patterns have been obtained.

Similar effects have been observed in the interactions of solitons with
impurities. The latter break the translational symmetry of the
unperturbed system and create an effective potential for scattering or
capture of the solitons \cite{kiv1}. Resonance effects in the
kink-impurity interaction have been investigated in \cite
{kiv2,fei1,fei2}. It has been shown in particular that kinks can be
reflected by an attractive impurity via a "two-bounce" resonance
mechanism, analogous to that of kink-antikink interaction involving the
excitation and deexcitation of a localized impurity mode \cite{fei1},
or an impurity and a shape mode \cite{fei2}. Scattering of NLS solitons
from point defects has been studied in \cite{kiv3,kiv4,cao} involving a
variety of nonresonant outcomes.

A problem of considerable theoretical and practical importance is the
interaction of solitons with extended defects
\cite{shar,ting,frau,kalb}. Trapping of solitons in potential wells and
nonclassical behavior for kinetic energies close to the height of
potential step have been obtained in \cite{kalb}, but no resonance
phenomena have been observed. The investigations however have been
restricted to potential widths comparable to the soliton width. In the
present work we investigate in detail the dynamics of bright NLS
solitons impinging on potential wells with variable shapes. For fixed
initial velocities slightly below the threshold for transmission, the
increase of the width of the well yields alternating regions of capture
and transmission, and occasionally - narrow reflection windows. The
regions of transmission, capture and reflection follow a remarkable
periodicity. The observed effects are explained by excitation and a
following resonant deexcitation of amplitude (shape) oscilations of the
soliton at the boundaries of the well. These oscillations are not
associated with true internal shape modes of the soliton, but with
dispersive radiation modes.

As a model in our numerical simulations we used the discrete nonlinear
Schr\"{o}dinger equation which describes the dynamics of nonlinear
Bose-type excitations in atomic or molecular chains in the presence of
defects which change locally the energy:

\begin{eqnarray}
 i\frac{\partial\alpha_{n}}{\partial t} =
 -(\alpha_{n+1} + \alpha_{n-1} - 2\alpha_{n})
 - 2\mid\alpha_{n}\mid^{2}\alpha_{n} + d_n \alpha_{n} \, .
\end{eqnarray}

In the continuum limit corresponding to wide solitons (compared to the
lattice constant) it turns into a perturbed NLS equation:

\begin{equation}
 i\frac{\partial \alpha}{\partial t} +
 \frac{\partial^{2}\alpha}{\partial x^{2}} +
 2\mid\alpha\mid^{2}\alpha = d(x)\alpha \, .
\end{equation}

For $d(x)\equiv 0$ (2) possesses a fundamental bright soliton
solution:

\begin{eqnarray}
\alpha(x,t) = \frac{1}{L}{\rm sech} \frac{x-vt-x_0}{L}e^{\textstyle
i(vx/2-\omega t)} \, , \quad \omega = \frac{v^2}{4} - \frac{1}{L^{2}}
\, ,
\end{eqnarray}
where $L$ and $v$ are the width and the velocity of the soliton.

It is well known, that for $d_n=0$ (1) is a nonintegrable
discretization of the completely integrable continuum NLS equation. For
sufficiently wide solitons however, the discreteness-induced effects
are negligible, and the solution (3) is stable on ideal discrete
lattices. We checked numerically the stability of (3) with $L\ge 4$ for
very long time intervals. Thus (3) was input as initial condition in
the simulations, placed 50 sites away from the defect region to avoid
radiation losses due to overlapping. A predictor-corrector method
\cite{sham} was used, periodic boundary conditions and chains much
longer than the defect region in order to eliminate boundary effects.
The accuracy of the calculations was controlled through the
conservation of the norm (number of particles), which is better than
$10^{-6}$.

The total energy associated with the solution (3) on an ideal lattice is:

\begin{equation}
E_s=\int_{-\infty} ^{\infty}(\mid\frac{\partial\alpha}{\partial
x}\mid^{2} - \mid\alpha\mid^{4})\,dx = \frac{v^2}{2L} - \frac{2}{3L^3}
\, ,
\end{equation}
where the first term describes the kinetic energy of the free
quasiparticles, and the second term - the nonlinear energy responsible
for the formation of the soliton. The evolutionary pattern depends in
general on the interplay between these two energies and the energy of
interaction with the defects $E_d$

\begin{equation}
E_d= \int_{-\infty} ^{\infty} d(x) \mid\alpha\mid^{2}\,dx
\end{equation}

The effects studied below correspond to the most interesting case of
slow solitons with kinetic energy of the order of the energy of
interaction with the defect and large nonlinear energy:
\[E_{kin} \sim |E_d| \ll |E_{int}| \]
or equivalently:
\begin{equation}
v^2L^2 \sim 2|d|L \ll 1 \, .
\end{equation}
The large nonlinear energy is required to preserve the integrity of the
soliton during the interaction.

Scattering of bright NLS solitons from single point defects has
been studied in detail in \cite{kiv3,kiv4,cao}. The corresponding
interaction energy when the soliton is on top of the defect is:

\begin{equation}
E_d= d\int_{-\infty} ^{\infty}\delta(x) \mid\alpha\mid^{2}\,dx = \frac{d}{L^2}
\end{equation}
When $E_{kin} \gg |E_d|$, the solitons are not influenced considerably
by the defect, and for $E_{kin} \ll |E_d|$ the solitons are reflected
even by an attractive defect. The possible outcomes in the case of slow
solitons ($E_{kin} \sim |E_d|$) and moderate defect strengths are
transmission or capture. No resonance reflection windows have been
obtained. A natural question arises as to what happens when the defect
spreads over several lattice sites. The energy of interaction with $N$
consecutive defects when the soliton is in the middle of the defect
region is:

\begin{equation}
E_d= d\int_{-N/2} ^{N/2}\mid\alpha\mid^{2}\,dx =
\frac{2d}{L}\tanh\frac{N}{2L}
\end{equation}

One can expect, that for a small number of defects, ($N \le L$), the
evolution should be similar to this of a soliton interacting with a
single defect with $N$-times greater strength. This turns out to be
true only for kinetic energies much smaller than the interaction
energy. In the case of comparable energies, the delocalization of the
defect (which decreases the interaction energy) can change the
evolution from capture to transmission (Fig.~1). It is worth noting
that the sharper the defect - the stronger the radiation accompanying
the interaction.

The focus of the present study lies in the interaction of NLS solitons
with potential wells considerably wider than the soliton width. We
modelled rectangular potential wells by $N$ equal consecutive defects
with $d=-0.007$ and used solitons in the form (3) with widths $L=5.75$,
which are stable on an ideal lattice. For initial velocities $v< 0.04$
the solitons are trapped inside the well and for $v>0.06$ they pass
through it and escape to infinity. For initial velocities in the
intermediate region, the outcome pattern as a function of the width of
the well exhibits periodically alternating regions of transmission and
capture, and occasionally, at the boundaries between them - narrow
reflection windows. This is shown on Fig.~2 where we have plotted the
final velocity of the soliton as a function of the width of the well
for different values of the initial velocity. When trapped, the soliton
oscillates back and forth inside the well with zero average velocity
which we have plotted as final. The sharp minima with negative final
velocity ($v_f<0$) correspond to reflection. These reflection "windows"
are extremely sensitive to the initial velocity. The relative widths of
the regions of transmission and capture depend on the initial velocity
and can be quite different, but the period of repeat remains constant
and depends weakly on the velocity.

Fig.~3 illustrates the evolutionary patterns corresponding to capture,
transmission and reflection. It is clearly seen, that amplitude (shape)
oscillations are excited when the soliton enters the potential well and
persist while the soliton is inside the well. Whenever the soliton
leaves the defect region, the shape oscillations are almost totally
extinguished. This brings forward the explanation of the observed
resonance phenomena: when the soliton reaches the defect region it
interacts inelastically with the sharp boundary and looses part of its
kinetic energy exciting additional modes. These can be internal shape
modes of the soliton and/or dispersive modes (radiation). We can
distinguish between them by examining the frequency of the shape
oscillations. The frequencies of the true internal shape modes can be
obtained by adding small perturbation $\alpha_{1}(x,t) =
\varphi_{1}(x)e^{\textstyle -i\Omega t}$ to the unperturb solution (3)
and solving the linearized Schr\"{o}dinger equation for it:

\begin{equation}
 \frac{\partial^{2}\varphi_{1}}{\partial x^{2}} +
 (\Omega + \frac{4}{L^2 \cosh^2(x/L)}) \varphi_{1} = 0 \, .
\end{equation}
The frequencies of the shape mode determined from (9) are $\Omega_1 = -
0.0735$ and $\Omega_2 = - 0.00094$. The period of the amplitude
oscillations obtained from the numerical data on Fig.~3 is T=208 which
yields a frequency of $\Omega = -2\pi/T = 0.030$. It practically
coincides with the internal frequency of the unperturbed soliton (3)
$\omega = -0.0296$. Shape oscillations with the internal soliton
frequency have been reported in \cite{Chbat} and are explained by
interference of the soliton with radiation modes with twice the soliton
frequency. This shows that similarly to the case of collision of vector
solitons \cite{Yang1}, the amplitude oscillations which we observe are
due to radiation, emitted during the inelastic interaction of the
soliton with the boundary, and not to true internal shape modes. The
soliton crosses the defect region accompanied by the radiation and this
configuration is stable and weakly damped. When the soliton meets the
second boundary, different outcomes are possible depending on the
timing. In the general case, as the initial velocities are very small,
the reduced kinetic energy of the soliton is not sufficient to overcome
the potential barrier of the second boundary, the soliton is reflected
from it and eventually gets trapped. However, the interaction of the
soliton with the boundary is phase-sensitive and if the time for which
the soliton crosses the defect region is commensurate with the period
of the shape oscillations, the inelastic interaction with the second
boundary can extinguish the shape oscillations, adding their energy
back to the kinetic energy of the translational motion and allowing the
soliton to overcome the barrier and escape to infinity which leads to
transmission. The higher the initial velocity of the soliton - the
wider the transmission regions (Fig.~2). In some cases the resonant
condition for escape is achieved after the soliton has crossed the
defect region twice - in the forward and backward directions. This
yields the observed narrow reflection windows which are analogous to
the three-bounce resonances observed in \cite{camp2}. Due to the
radiation losses accompanying the propagation of the soliton inside the
well, these higher-order resonances are very sharp, extremely sensitive
to the initial velocity and difficult to observe.

The interaction energy when the soliton is at the boundary of the
defect region is:
\begin{equation}
E_d= d\int_0 ^{N} \mid\alpha\mid^{2}\,dx = \frac{d}{L} \tanh
\frac{N}{L}
\end{equation}
For narrow potential wells, comparable to the soliton width, the
interaction energy is smaller, leading to weaker shape
oscillations and a narrower region of trapping. For sufficiently
wide potential wells ($N\gg L$, $\tanh (N/L) \sim 1$), when the
soliton does not "feel" the second boundary, the interaction
energy and the excited shape oscillations are constant, which
results in the observed periodic outcome pattern. For very wide
potential wells, due to the radiation losses accompanying the
oscillating soliton, the regions of transmission get narrower and
eventually close down.

An increase of the depth of the well leads to wider regions of trapping
and narrower regions of transmission (Fig.~4), while the total period
remains unchanged. The inelastic interaction of the soliton with the
boundary is stronger in this case and a larger portion of the kinetic
energy of the soliton is transformed into radiation. A more exact
resonance condition is required at the second boundary for the escape
of the soliton, which yields narrower regions of transmission.

Contrarily, the change of the shape of the potential well from
rectangular to trapezoid leads to wider transmission regions and
narrower regions of capture (Fig.~5). The potential in this case is
smoother and the perturbation it induces is weaker. Hence a smaller
portion of the kinetic energy is transformed into radiation and the
resonance condition at the second boundary is more relaxed.

We also checked the dependence of the evolutionary pattern on the
initial position of the soliton with respect to the boundary of the
defect region. For an initial soliton in the form (3) and a fixed
velocity $v=0.05$ we obtain a threshold initial distance of 15 lattice
sites, above which the soliton passes through the defect region and
below it gets trapped. This can be explained by the radiation losses
when the initial unperturbed soliton overlaps with the defect region.
The chaotic behavior of the outcome with the initial soliton position
obtained in \cite{kalb} can be attributed to the different type of
coupling between the soliton and the shape mode.

\begin{acknowledgments}
This work is supported in part by the National Science Foundation of
Bulgaria under Grant No. F911.
\end{acknowledgments}

\begin{figure}
\resizebox{2.5in}{7.in} {\includegraphics{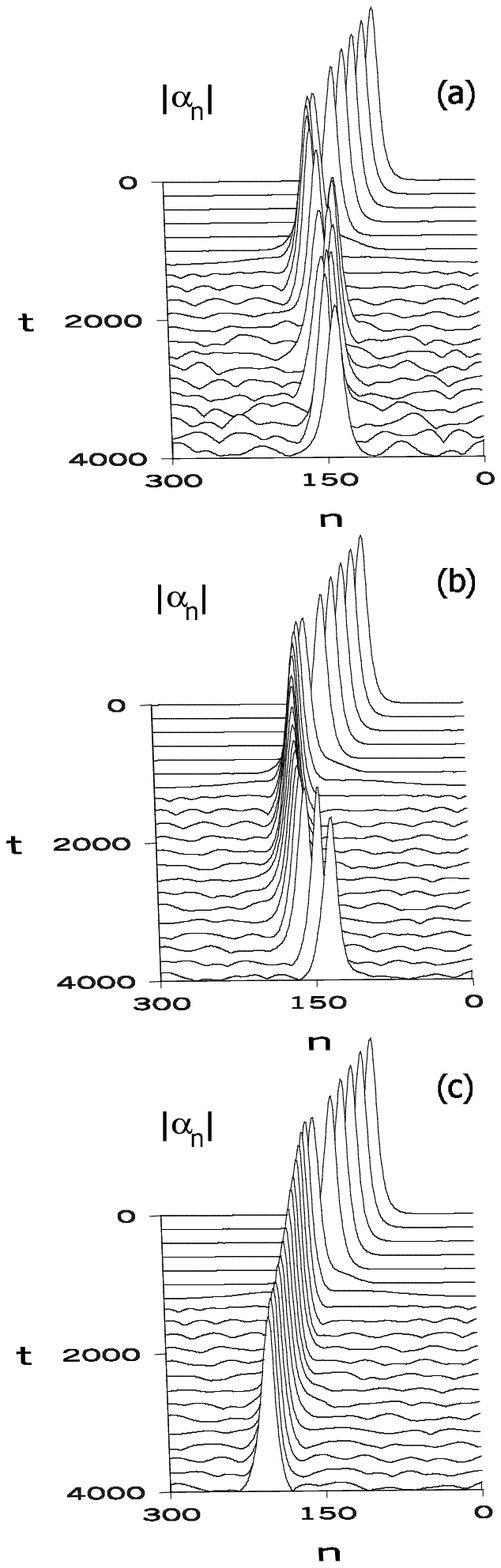}} \caption{\label{fig.1}
Soliton interaction with a small number of impurities for $v=0.05$. (a)
$N=1$, $d=-0.035$; (b) $N=2$, $d=-0.0175$; (c) $N=3$, $d=-0.0117$. (a)
and (b) correspond to capture and (c) - to transmission.}
\end{figure}

\begin{figure}
\resizebox{5.in}{!} {\includegraphics{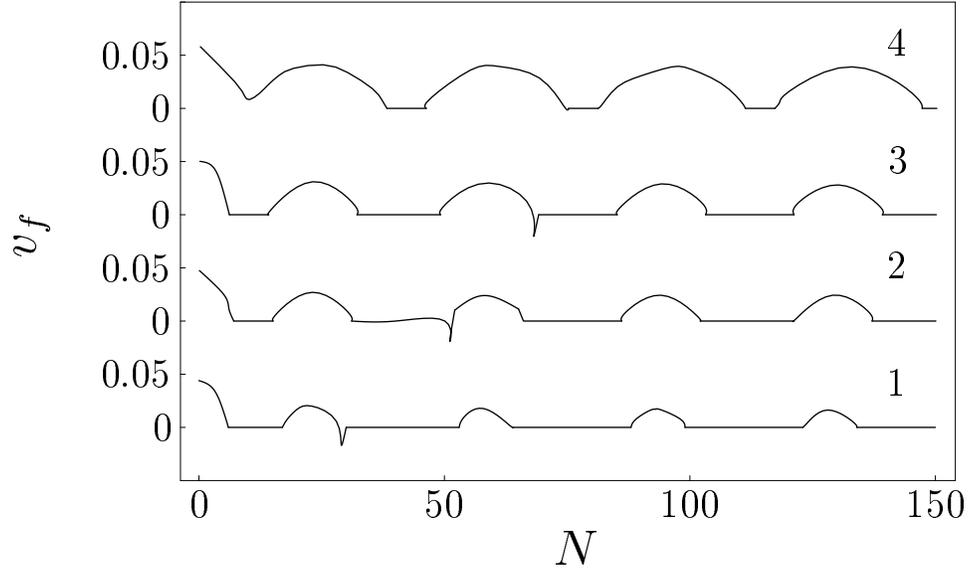}} \caption{\label{fig.2}
Final soliton velocity $v_f$ for $d=-0.007$ and different initial
velocities $v$. Curves $1$ to $4$ correspond to $v=0.0440$; $0.0476$;
$0.0502$ and $0.0580$ respectively.}
\end{figure}

\begin{figure}
\resizebox{5.in}{!} {\includegraphics{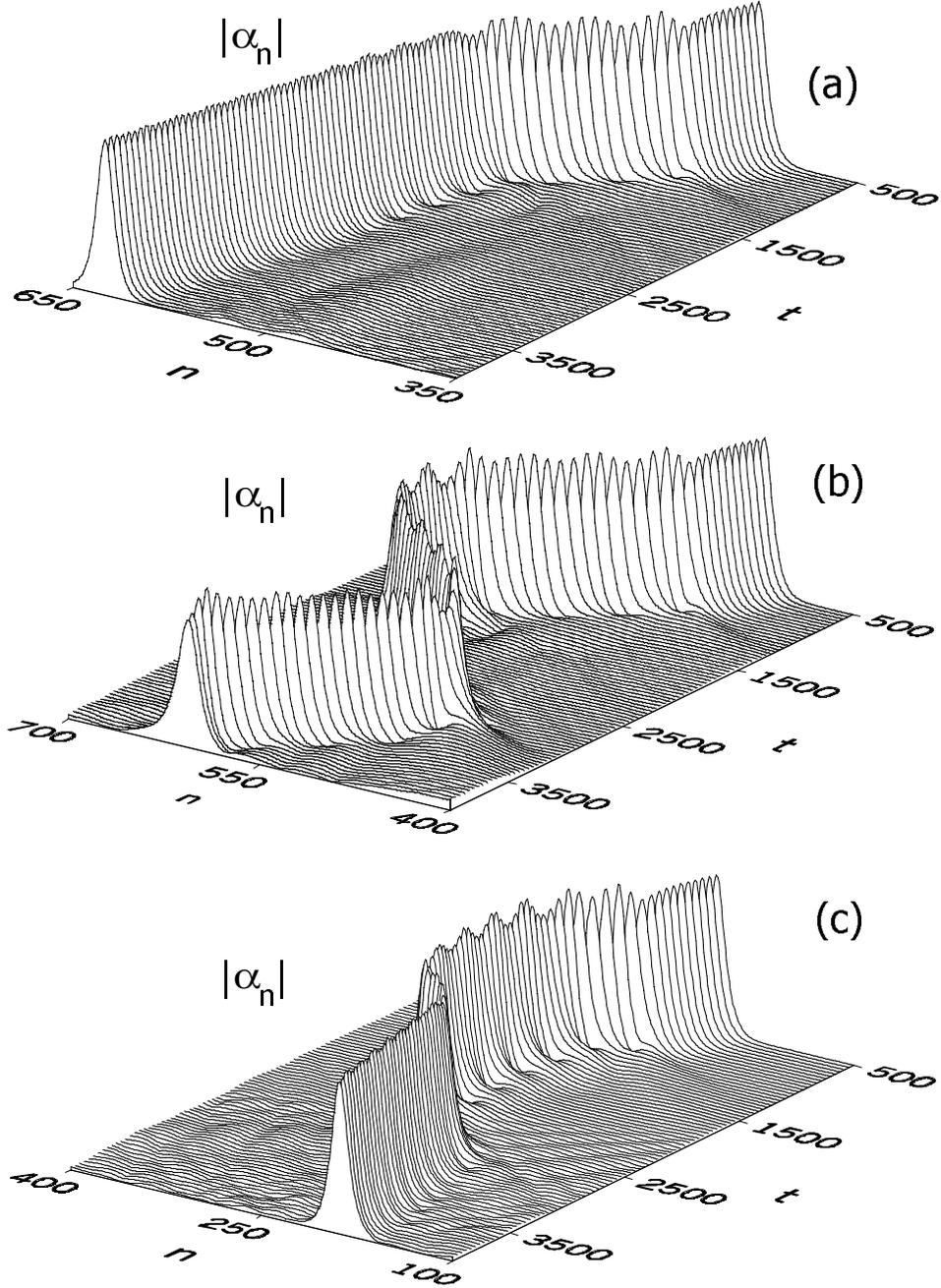}} \caption{\label{fig.3}
Outcome patterns for $v=0.05$, $d=-0.007$ and different number $N$ of
defects. (a) $N=95$  - transmission, (b) $N=110$ - trapping and (c)
$N=33$ - reflection. Only a portion of the chain around the defect
region is presented, while the the total length of the chain exceeds 10
potential widths (to avoid boundary effects).}
\end{figure}

\begin{figure}
\resizebox{5.in}{!} {\includegraphics{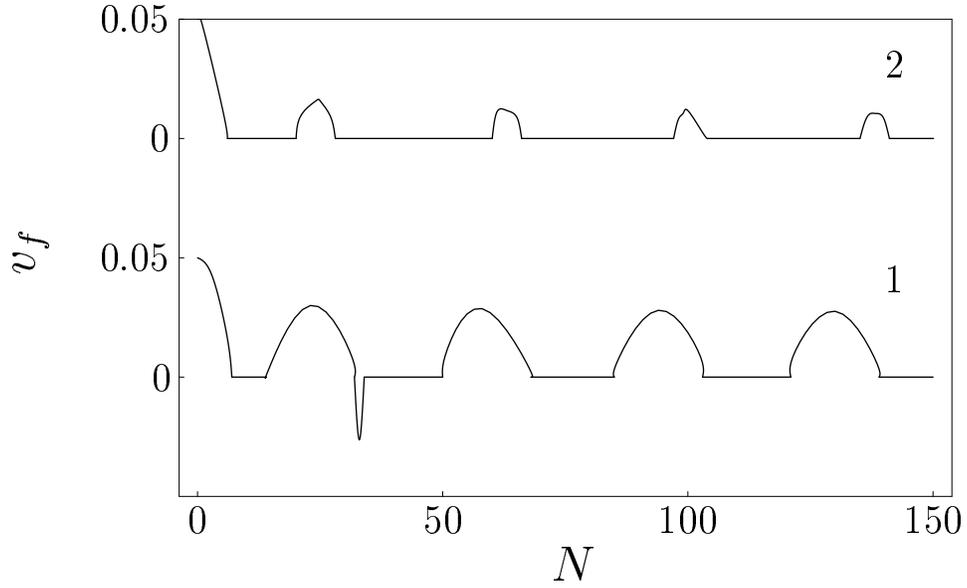}} \caption{\label{fig.4}
Final soliton velocity $v_f$ for $v=0.05$ and different depths of the
potential well; curve $1$ - $d=-0.007$ and curve $2$ - $d=-0.008$.}
\end{figure}

\begin{figure}
\resizebox{5.in}{!} {\includegraphics{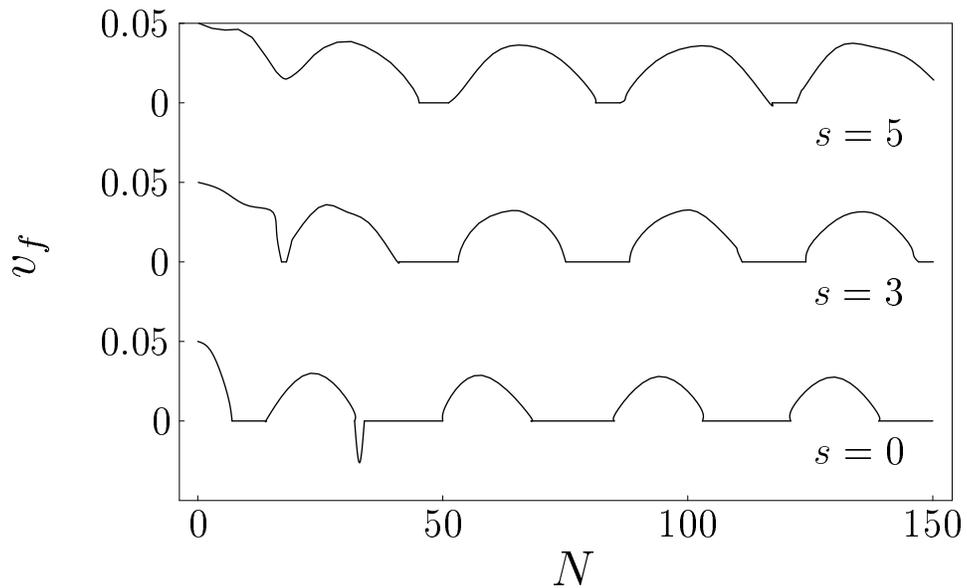}} \caption{\label{fig.5}
Final soliton velocity $v_f$ for $v=0.05$, $d=-0.007$ and trapezoidal
potential wells with different slope ($s$ is the extension of the
slope).}
\end{figure}

\end{document}